# Choriocapillaris Flow Signal Impairment in Sorsby Fundus Dystrophy

Kristina Hess[1,2], Kristin Raming[1,2], Martin Gliem[3], Peter Charbel Issa[4,5], Philipp Herrmann[1,2], Frank G. Holz[1,2], Maximilian Pfau[1,6]

1. Department of Ophthalmology, University of Bonn, Bonn, Germany
2. Center for Rare Diseases Bonn, University of Bonn, Bonn, Germany
3. Boehringer Ingelheim International GmbH, Ingelheim am Rhein, Germany
4. Oxford Eye Hospital, Oxford University Hospitals NHS Foundation Trust, Oxford, United Kingdom
5. Nuffield Laboratory of Ophthalmology, Department of Clinical Neurosciences, University of Oxford, Oxford, United Kingdom
6. Ophthalmic Genetics and Visual Function Branch, National Eye Institute, Bethesda, MD, USA

| | |
|---|---|
| **Short Title:** | Choriocapillaris flow impairment in SFD |
| **Keywords:** | Sorsby Fundus Dystrophy, Choriocapillaris, Bruch Membrane, Optical coherence tomography angiography, Outcome measures, Clinical trials |
| *Words:* | *2384* |
| *Figures:* | *5* |
| *Tables:* | *0* |
| *Supplementary figures:* | *2* |

**\*Corresponding author:**
Kristina Hess, MD
University Hospitals Bonn
Department of Ophthalmology
Ernst-Abbe-Strasse 2
53111 Bonn
GERMANY
Tel:     +49 228-287-15505
Fax:     +49 228-287-14817
Mail: Kristina.Hess@ukbonn.de

# ABSTRACT

**Purpose**

To quantify choriocapillaris flow alterations in early Sorsby Fundus Dystrophy (SFD) and to investigate the relationship of choriocapillaris flow with the choroidal and outer retinal microstructure.

**Methods**

In this prospective case-control study, 18 eyes of 11 patients with early SFD and 32 eyes of 32 controls without ocular pathology underwent multimodal imaging including spectral-domain optical coherence tomography (OCT) followed by deep-learning-based layer segmentation. OCT-angiography (OCT-A) was performed to quantify choriocapillaris flow signal voids (FVs). Differences in choriocapillaris flow (FVs) area percentage between SFD patients and controls were determined and a structure-function correlation with outer retinal layer thicknesses were analyzed based on mixed model analysis.

**Results**

SFD patients exhibited a significantly greater choriocapillaris FVs area percentage than controls (estimate [95% CI] 32.05% [24.31–39.80] vs. 23.36 % [20.64–26.09], P<0.001), even when adjusting for age. Choroidal thickness was a structural OCT surrogate of the choriocapillaris FVs area percentage (-0.82% per 100µm, P=0.017), whereas retinal-pigment-epithelium-drusen-complex thickness was not informative regarding choriocapillaris FVs (P=0.932). The choriocapillaris FVs area percentage was associated with an altered microstructure of the overlying photoreceptors (outer-segments, inner-segments and outer-nuclear-layer thinning of -0.31 µm, -0.12 µm and -0.47µm per %FVs, respectively, P<0.001).

**Conclusions**

Patients with early SFD exhibit pronounced abnormalities of choriocapillaris flow signal in OCT-A, which are not limited to areas of sub-RPE deposits seen in OCT imaging. Thus, analysis of the choriocapillaris flow may enable clinical trials at earlier disease stages in SFD and possibly in mimicking diseases with an impaired Bruch membrane including age-related macular degeneration.

**Précis:**

Choriocapillaris flow alterations are present in early stages of Sorsby Fundus Dystrophy and precede structural alterations of the outer retina and retinal pigment epithelium. It is therefore suitable to quantify early disease progression which might be relevant for emerging interventional trials.

## INTRODUCTION

Bruch's Membrane (BrM) is a pentalayer of extracellular components, located between choriocapillaris -.,and retinal pigment epithelium (RPE). It constitutes a key interchange barrier for nutrients and waste between the systemic circulation and neuroretina.[1] Alterations of BrM have been identified in a variety of different disease entities, including age-related macular degeneration (AMD), which is the leading cause of blindness in industrialized countries[2] and - more rarely - primary BrM diseases such as Pseudoxanthoma elasticum and Late-onset retinal degeneration.[3–5]

Sorsby Fundus Dystrophy (SFD, OMIM 136900) represents one of the primary BrM diseases an autosomal-dominant inherited retinal disease, which is caused by a mutation in the *TIMP3*-gene coding for the tissue inhibitor of metalloproteinase 3 (TIMP-3) protein.[4] Physiologically, TIMP-3 binds to a number of extracellular membrane-type matrix metalloproteinases (MMPs) which play an important role for the homeostasis and turnover of extracellular matrix components.[6,7] TIMP3 can be found in BrM after secretion from the RPE. Mutant TIMP-3 protein accumulates at the level of BrM, resulting in thickening and, thus, assumed to cause a diffusion barrier impairing the interchange across the choriocapillaris-BrM-RPE-complex.[1,8]

For disease monitoring, dark adaptometry has been successfully used as a functional measure of the impaired trafficking across an altered BrM in SFD,[9] and likewise in AMD.[10,11] However, psychophysical testing is relatively time-consuming. Regarding imaging biomarkers, quantification of BrM-RPE separation has been proposed in the context of SFD.[12] But the application of this metric as outcome measure would systematically exclude early-stage SFD patients from clinical trials, due to the absence of visible BrM-RPE separation on optical coherence tomography, although these patients would hypothetically benefit the most from a therapeutic intervention in the long run.

Therefore, the aim of this study was to investigate the impact of a diseased BrM on choriocapillaris flow quantified by optical coherence tomography angiography (OCTA) as a potential marker for early disease stages in SFD. Moreover, we aimed to identify correlates of reduced choriocapillaris flow on widely available structural spectral-domain optical coherence tomography (OCT) and to identify structural associations of reduced choriocapillaris flow and overlying photoreceptor microstructure.

# METHODS

## Patients

Patients with SFD were recruited between June 2019 and May 2020 from a dedicated clinic for inherited retinal diseases of the Department of Ophthalmology, University Hospital Bonn, Germany. This prospective single-center case-control study was in adherence to the Declaration of Helsinki. Institutional review board approval (Ethikkommission, Medizinische Fakultät, Rheinische Friedrich-Wilhelms Universität Bonn, Germany) and patients' informed consent was obtained.

Inclusion criteria were the clinical diagnosis of SFD based on funduscopy, optical coherence tomography and fundus autofluorescence, in conjunction with the presence of at least one pathogenic mutation in the *TIMP3* gene. Exclusion criteria were previous or current anti-VEGF treatment in the study eye and areas of atrophy larger than one optic nerve disc area. Further exclusion criteria were additional retinal pathologies not associated with SFD, refractive errors > ±3 dpt or opacities of the ocular media. Healthy eyes from age-similar subjects served as controls.

## Image Acquisition

All subjects and controls underwent a complete ophthalmologic examination including refraction and BCVA testing, slit lamp examination and – after pupil dilation with 0.5% tropicamide and 2.5% phenylephrine – ophthalmoscopy as well as an extensive imaging protocol. This included spectral-domain OCT (30° x 20° volume scan with 121 B-scans), short-wavelength fundus autofluorescence (FAF), near-infrared reflectance (NIR) images (all acquired with a Spectralis HRA-OCT 2, Heidelberg Engineering, Heidelberg,

Germany) and color fundus photography (Zeiss Visucam, Oberkochen, Germany or Eidon confocal fundus camera, Centervue, Padua, Italy).

OCTA was performed using a swept-source OCT device (6x6 mm OCTA scan, PLEX Elite 9000; CarlZeiss Meditec AG, Jena, Germany).

**Processing of OCT and OCTA Images**

For OCTA images, the automated segmentation algorithm of the device was initially applied, followed by manual correction of the BrM ("RPE fit") segmentation as needed. The choriocapillaris slab ranging from +29 µm to +49 µm below "RPE fit" was exported with removal of decorrelation tail artefacts.[13]

The SD-OCT data were segmented using a customized, previously validated deep-learning pipeline.[14] Thickness maps for the choroid, RPE drusen complex (RPEDC,[15,16] ranging from BrM to the upper boundary of the RPE or subretinal drusenoid deposits), photoreceptor outer segments (OS) and inner segments (IS) as well as the outer nuclear layer (ONL) were generated. The thickness of these layers were extracted for each ETDRS-grid subfield centered to the fovea.

**OCTA Image Analysis**

Choriocapillaris images with artifacts (e.g., due to pronounced unstable fixation) or marked shadowing (e.g., floaters) were excluded from the analysis. First, all images were imported into FIJI (Version 2.0.0, National Institutes of Health, Bethesda, MD, USA). We performed compensation of shadowing artifacts as previously descirbed.[17,18] Briefly, the CC-slab image was downscaled the angiography image to 512 x 512 pixels from originally 1024 x 1024 pixels. Then the OCTA images were multiplied with the inverted, smoothed structural image (Gaussian smoothing with a sigma of 3 pixels) to compensate for shadowing effects.

Subsequently, automatic local thresholding was performed with the Phansalkar method using a radius of 15 pixels to extract flow signal void. Last, an ETDRS-grid was centered to the fovea to extract the area percentage of flow signal voids (FVs) of the outer, inner and central ETDRS subfields.

The CC directly under large superficial retinal vessels, was excluded from the analysis as previously described to exclude severe shadowing or projection artifacts.[19]

Therefore, the structural ORCC image was imported to FIJI and the MaxEntropy threshold was applied to visualize only the secondary and tertiary superficial retinal vessels (of first and second order.(Figure 1) The identified areas were excluded from the analysis.

ETDRS subfields with any atrophic areas as seen on OCT-imaging and FAF were excluded from the analysis.

**Statistical Analyses**

Statistical analysis was performed using the software environment R (version 3.2.0; R Foundation for Statistical Computing, Vienna, Austria). Continuous variables were described by using the mean and standard deviation and categorical variables were analyzed using frequency tables. A P-value < 0.05 was considered significant. All P values were Bonferroni corrected. Mixed-effects models with patient ID as random effects term were used to account for the hierarchical nature of the data (eye nested in patient as random effects term). First, we investigated the association of age and ETDRS subfield location as explanatory variables (i.e., fixed effects) with the area percentage of FVs as independent variable. Likelihood ratio tests (LRTs) were applied to test for significance. Subsequently, we examined the association of structural indicators (choroidal thickness and RPEDC thickness) with FVs as independent variable. Last, the association of FVs (this time as dependent variable) with photoreceptor laminae integrity (e.g., ONL, IS, OS thickness as dependent variable) was examined.

## RESULTS

### Cohort Characteristics

Eighteen eyes from 11 patients (47.3±11.9 years, range 28-61 years, 7 female) with a median [IQR] best-corrected visual acuity of 0.0 logMAR [-0.01; 0.01] and 32 eyes from 32 age-similar controls (52.4 ±19.4 years, range 21-82 years, 13 female) were included in this study. In 3 eyes, ETDRS subfields were excluded from the analysis due to atrophy, leading to a total of 5 excluded subfields.

### Choriocapillaris FVs

According to mixed-model analysis, the choriocapillaris FVs area percentage was significantly associated with the subgroup (patients versus controls, univariate estimate [95% CI] +8.69 % [3.67; 13.71%]) as well as with age (+0.19 %/year [0.05; 0.33%], Figure 2). The estimates were similar for the multivariable analysis with a consistent, significant effect of the subgroup (patients versus controls, estimate [95% CI] +9.74 % [5.36; 14.12%]) as well as of age (+0.22 %/year [0.11; 0.34]). Neither the ETDRS subfield (Figure S1), nor an interaction term between age and subgroup (i.e., hypothesizing that patients exhibit a more severe age-dependent increase of FVs) exhibited a significant association with the choriocapillaris FVs area percentage.

### Structural Indicators of Choriocapillaris Flow Signal Voids

Choroidal thickness was significantly lower in patients (218.33 ± 91.40 μm) compared to controls (282.35 ± 90.02 μm, P<0.001, Figure 3A). Moreover, choroidal thickness showed a similar spatial pattern among all subjects (Supplementary Figure 2).

Choroidal thickness and the choriocapillaris FVs area percentage exhibited a significant, linear association (univariate estimate [95% CI] -0.01 % / µm [-0.02; 0.00], P=0.017, Figure 3B). Square -root or logarithmic transformation of choroidal thickness did not improve the model fit.

In contrast, the RPEDC thickness was not significantly associated with the choriocapillaris FVs area percentage (Figure 3D).

**Association with overlying Photoreceptor Integrity**

The three SD-OCT laminae corresponding to the retinal photoreceptors were all inversely associated with the underlying choriocapillaris FVs area percentage (i.e., thinning of these layers was evident in eyes with a greater FVs area percentage). Specifically, OS thickness was associated with a univariate estimate of -0.31 μm per % [-0.36; -0.26]. IS thickness showed a comparable association with -0.12 μm per % [-0.14; -0.10].

For the ONL thickness, the correlation was weaker (-0.47 μm per %, [-0.66; -0.27], P<0.001) and not reaching significance for inner retinal thickness (-0.41 μm per %FVs, P=0.084) after Bonferroni correction (Figure 4).

Clinical examples of patients in different disease stages are shown in Figure 5.

## DISCUSSION

To the best of our knowledge, this is the first study investigating choriocapillaris alterations using OCTA in patients with SFD. Our findings demonstrate a significantly increased choriocapillaris FVs area percentage indicative of impaired choriocapillaris flow in patients with SFD. This observation was evident even when adjusting for age. Moreover, an increased choriocapillaris FVs area percentage is associated with alterations of the overlying photoreceptor microstructure. Our findings highlight that SFD-associated alterations of BrM result (directly or indirectly) in both, degeneration of the choriocapillaris and outer retina.

The increased choriocapillaris FVs area percentage was present in eyes with no or only a mild phenotypic manifestation on color fundus photography, indicating that choriocapillaris changes may represent an early structural marker for BrM abnormalities. Possibly, the previously observed multilobular hypofluorescence in indocyanine green angiography in the central retina of patients with SFD without visible fundus changes reflects the same underlying loss of choriocapillaris perfusion (cf. Figure 7 in reference 19).[20] In principle, this is compatible with the histopathologic description of extensive choriocapillaris loss in SFD,[21] however choriocapillaris density for clinically unremarkable retinal regions has not been reported to the best of our knowledge. In other BrM diseases such as age-related macular degeneration, choriocapillaris alterations have been demonstrated even in regions without overlying retinal atrophy.[22,23] In Stargardt disease and Pseudoxanthoma elasticum, characteristic choriocapillaris alterations on late phase fluorescein and indocyanine green angiography are pathognomonic, and helpful to establish the underlying diagnosis.[24,25]

Biologically, the choriocapillaris flow deficit in SFD is mostly likely a secondary phenomenon given that TIMP3 is primarily expressed by the RPE.[26] However, it is unclear, whether the choriocapillaris flow deficit is merely a response to an altered RPE, or whether an interchange barrier between the RPE and choriocapillaris drives subsequent choriocapillaris degeneration.[21,26,27]

In our study, choriocapillaris FVs were significantly correlated with choroidal thickness. An increased choriocapillaris FVs area percentage was pronounced for choroidal thicknesses of 200 μm or less. Accordingly, choroidal thickness may possibly serve as a surrogate for sub-RPE/ choriocapillaris perfusion. Previous studies in patients with BrM diseases including SFD and pseudoxanthoma elasticum indicated an inverse association between disease severity and choroidal thickness.[28,29] In contrast, RPEDC thickness as a measure of the amorphous deposits found between the basement membrane of the RPE and the inner collagenous layer of BrM was not associated with the choriocapillaris FVs area percentage. Estimating disease severity based on RPEDC thickening, which has been recently suggested for SFD and autosomal-dominant drusen,[12] may thus be unsuitable to quantify the overall disease severity in early stages. Moreover, choriocapillaris degeneration appears to be more widespread than (visible) alterations at the level of RPEDC.

Last, our findings indicate that an impaired choriocapillaris-BrM complex is associated with degeneration of the overlying photoreceptors. Specifically, we showed a negative correlation of the choriocapillaris FVs area percentage with the overlying ONL, IS and OS thicknesses. This thinning of the outer retinal laminae most likely represents downstream degeneration of photoreceptors due to RPE dysfunction and possibly chronic local deprivation of Vitamin A.[9,30]

Early choriocapillaris changes highlight that efficacy of therapeutic interventions may be subject to a time window. It was previously demonstrated in SFD, as well as other diseases of BrM such as AMD and Pseudoxanthoma elasticum, that oral intake of high dosages of Vitamin A may improve rod mediated dark adaptation.[9,31,32] Given the herein observed loss of choriocapillaris with age and the associated microstructural alterations of photoreceptors, earlier rather than later treatment may be warranted. Otherwise, the residual choriocapillaris density and photoreceptor degeneration may limit the therapeutic efficacy, and, thus, the visual outcome of a Vitamin A supplementation. Similar consideration would apply to emerging therapeutic strategies targeting BrM, including subthreshold laser therapy to induce RPE matrix metalloproteinases expression,[33] and Apolipoprotein A-I mimetics for lipid clearance,[34] too.

Limitations of this study include the relatively small sample size, which constitutes a common challenge in rare diseases. The variability of choroidal thickness in the normal population has been addressed by adjusting for age and exclusion of patients with higher refractive errors. The observed difference between SFD patients and controls must be considered as conservative estimates given that imaging at the same time in consideration of circadian changes may further improve the signal-to-noise ratio of our analysis. Quantitative analysis of choriocapillaris FVs is affected by the slab selection and thresholding/binarization approach. [35–37]  Here, we applied the preset slab definitions, however other slab definitions may possibly further accentuate differences between patients and controls. Choriocapillaris FVs data were summarized using the well-established ETDRS-grid. Other patterns with smaller regions-of-interest could potentially highlight more subtle relationships between outer retinal integrity and choriocapillaris.

 Further, the choriocapillaris FVs area percentage may be affected by shadowing, which is of importance in SFD due to clinical endpoint of extensive subretinal material and BRM/RPE changes. However, we used long-wavelength swept-source OCTA, performed signal compensation with the structural *en-face* slab and solely included early and intermediate stages of SFD. Moreover, the fact that we observed no association between RPEDC thickening and the choriocapillaris FVs area percentage is further indicative of the absence of shadowing artifacts.

Our findings indicate an association between changes in the BrM-RPE complex and rarefication of choriocapillaris flow, associated with a phenotype with carious similiarities to AMD. This adds evidence for interdependent roles within the choriocapillaris-BrM-RPE-complex as well as it underscores the importance of BrM and the choroid as pathogenic factors for disease development and progression of diseases involving BrM abnormalities, including AMD.

# Acknowledgements

a.  **Funding/Support:** This work was supported by BONFOR program from the University Bonn (grant 2019 1A-13 to KH), the German Research Foundation (DFG grant PF 950/1-1 to MP). PCI was supported by the National Institute for Health Research (NIHR) Oxford Biomedical Research Centre (BRC), Oxford, United Kingdom. The views expressed are those of the authors and not necessarily those of the NHS, the NIHR or the Department of Health. The sponsor and funding organization had no role in the design or conduct of this research.

b.  Financial disclosures:  The views expressed are those of the authors. The funding organization had no role in study design, data collection, analysis, or interpretation, or the writing of the report.

c.  **Financial Disclosures:** Carl Zeiss MedicTec has provided research material (PLEX Elite 9000) for the conduct of this study. Carl Zeiss MedicTec had no role in the design or conduct of the experiments.

   KH: Heidelberg Engineering: non-financial support, Carl Zeiss MedicTec AG: non-financial support, Optos: non-financial support

   KR: Heidelberg Engineering: non-financial support, Carl Zeiss MedicTec AG: non-financial support, Optos: non-financial support

   MG: Employee and equity owner of F. Hoffmann-La Roche, Basel, Switzerland.

   PCI: Consultant for Gyroscope and Inozyme, has received research support from Heidelberg Engineering and travel support from Bayer; and he has been a principle investigator on a number of commercial trials sponsored by NightStar, Acucela, Gyroscope, and Apellis.

   PH: Financial support for presentations held for Novartis GmbH, Bayer Health AG, Allergan-Pharm GmbH. Heidelberg Engineering: non-financial support, Carl Zeiss MedicTec AG: non-financial support, Optos: non-financial support

   FGH: Heidelberg Engineering: Grant, Personal fees, non-financial support, Novartis: Grant, personal fees, Bayer: Grant, personal fees, Genetech: Grant, personal fees, Acucela: Grant, personal fees, Boehringer Ingelheim: Personal fees, Alcon: Grant, personal fees, Allergan: Grant, personal fees, Optos: Grant, personal fees, non-financial support, Carl Zeiss MediTec AG: non-financial support

   MP: Heidelberg Engineering: non-financial support, Carl Zeiss MedicTec AG: non-financial support, Optos: non-financial support

d.  **Other Acknowledgements:** None

# FIGURE CAPTIONS

## Figure 1. Data processing and analysis steps in an exemplary eye

Flow (A) and structure (B) images from optical coherence tomography angiography (OCTA) were combined to compensate for signal attenuation (C) and binarized using the Phansalkar method (D). Large vessels were excluded from the analysis (marked in red, E).

For data analysis, an ETDRS-grid was superimposed (F) and each sector was analyzed regarding the percentage of flow void.

Spectral-domain optical coherence tomography (SD-OCT) images were segmented in a semi-automated manner.  RPEDC = Retinal pigment epithelial drusen complex

**Figure 2. Association of age and FVs in patients and controls.**

Patients with Sorsby Fundus Dystrophy (SFD) exhibited on average a significantly higher flow void percentage (+8.69% for SFD vs. controls). A positive association of FVs with age (0.21%/year) was evident for healthy and patients with Sorsby fundus dystrophy (SFD).

The age range of SFD patients is limited compared to the age range controls due to the time until diagnosis and due to increasing risk of neovascularization with age, which results in exclusion from the study.

**Figure 3. Association of choroidal thickness with choriocapillaris FVs**

In A, a significantly reduced choroidal thickness in patients with Sorsby Fundus Dystrophy (SFD) was evident despite the slightly younger mean age of patients (47.3 ± 11.9 years) compared to controls (52.4 ± 19.4 years, p=0.3).

A linear negative association of choroidal thickness and choriocapillaris FVs could be observed in B. SFD patients (red) exhibit overall lower choroidal thickness values and more FVs.

In patients with choroidal thicknesses below 200μm, flow void percentage increased markedly.

In C, an overall increased RPEDC-thickness and a wider range of RPEDC-thicknesses can be shown for SFD patients. RPEDC-thickness exhibited mild positive association with choriocapillaris FVs, not reaching the significance level. Compared to the associations FVs with age (Figure 3) and choroidal thickness (Figure 3, A&B), RPEDC-thickness exhibited the weakest association with FVs.

**Figure 4. Association of choriocapillaris flow voids and structural alterations.**

All three layers representing photoreceptor integrity show marked thinning with increasing flow voids. The most pronounced thinning is evident for the photoreceptor outer segments (OS) shown in A and the photoreceptor inner segments (IS) in B. The outer nuclear layer (ONL) in C shows a weaker association.

**Figure 5. Progression of choriocapillaris flow deficits in different disease stages**

The upper row shows a healthy eye on optical coherence tomography (OCT, first column), fundus photography (FP, second column), fundus autofluorescence (BAF, third column). In the fourth column, the unprocessed choriocapillaris OCT Angiography (OCTA) image is shown and the binarized/thresholded image is shown in the $5^{th}$ column.

Row one shows a patient with Sorsby Fundus Dystrophy (SFD) with unremarkable OCT, FP and FAF. However, the binarized/thresholded image shows slightly increased flow voids (black areas) compare to the healthy eye.

In row two, an SFD patient with reticular pseudodrusen (RPD) on OCT imaging (first column) exhibits a characteristic pattern of RPD on FP and FAF. OCTA shows an irregular pattern of choriocapillaris flow with patchy black areas in the central macular (yellow circle)

The third row shows a patient with pronounced alterations on OCT, FP, FAF including soft drusen, RPD and beginning atrophy. OCTA of the choriocapillaris layer reveals marked flow voids, visible as dark areas (highlighted by yellow arrows) and more subtle, smaller flow void areas (yellow arrowheads) as well as a more irregular pattern of flow.

| CC- Flow | CC-Structure | Signal compensation | Binarization | Exclusion of large vessels |
|---|---|---|---|---|

Processing

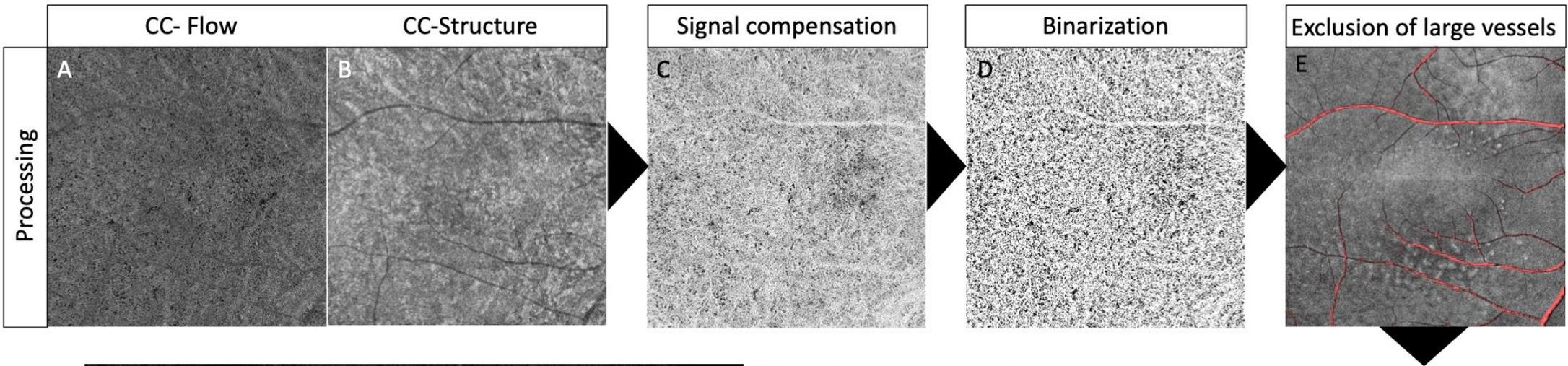

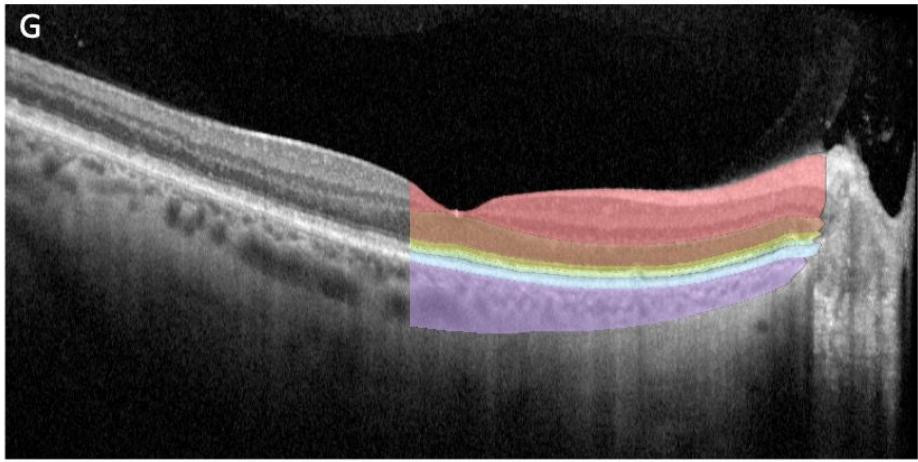

Inner retina
Outer nuclear layer
Inner segments
Outer segments
RPEDC
Choroid

Analysis

Flow voids %

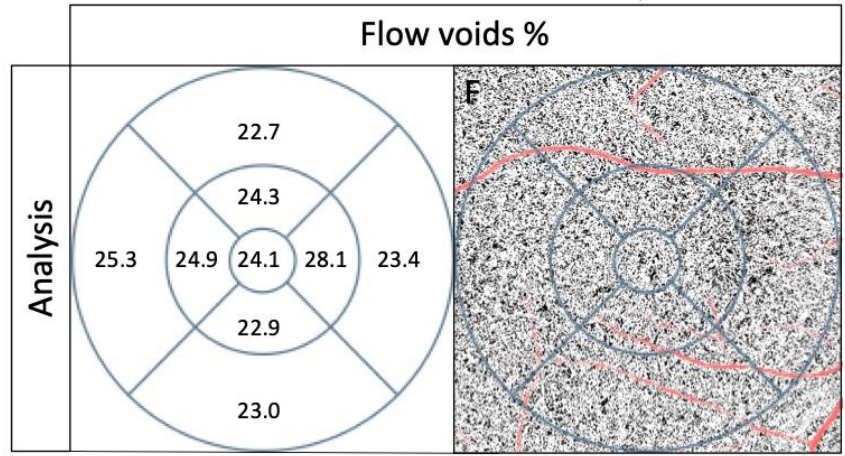

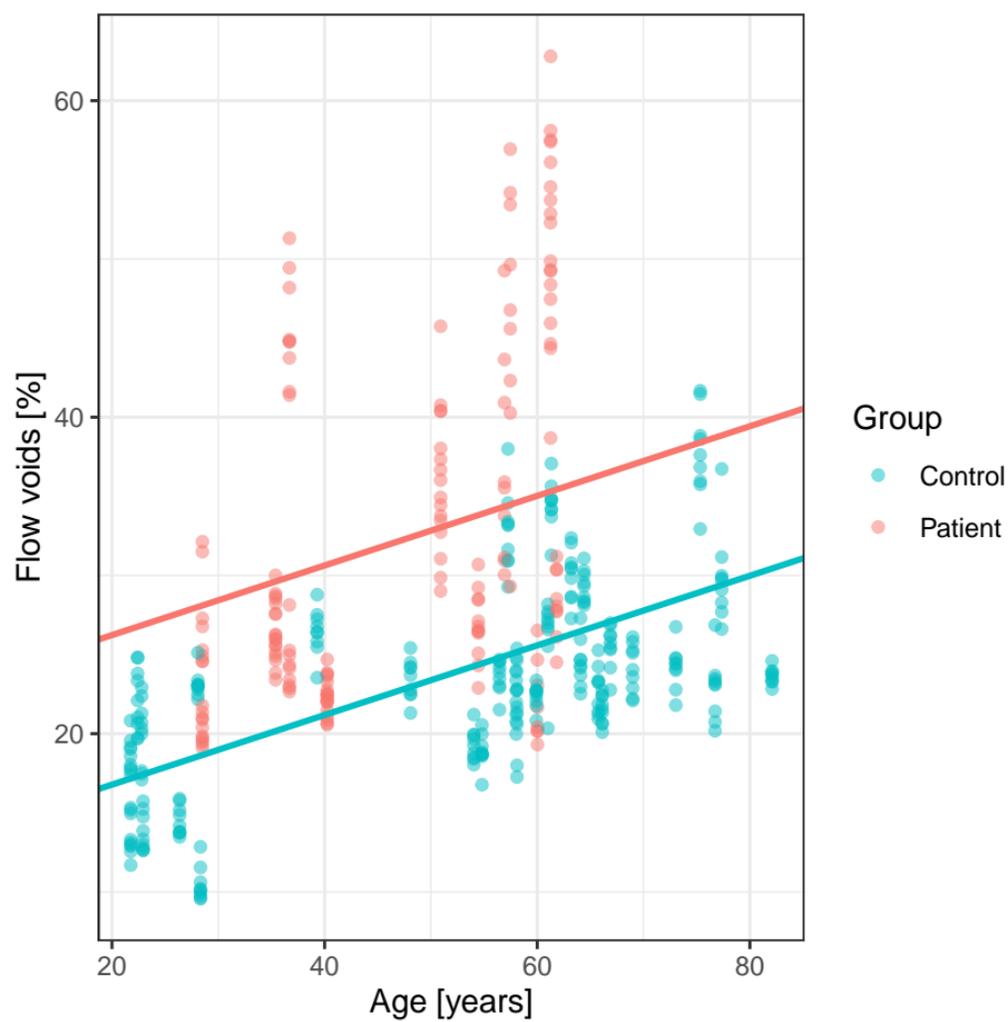

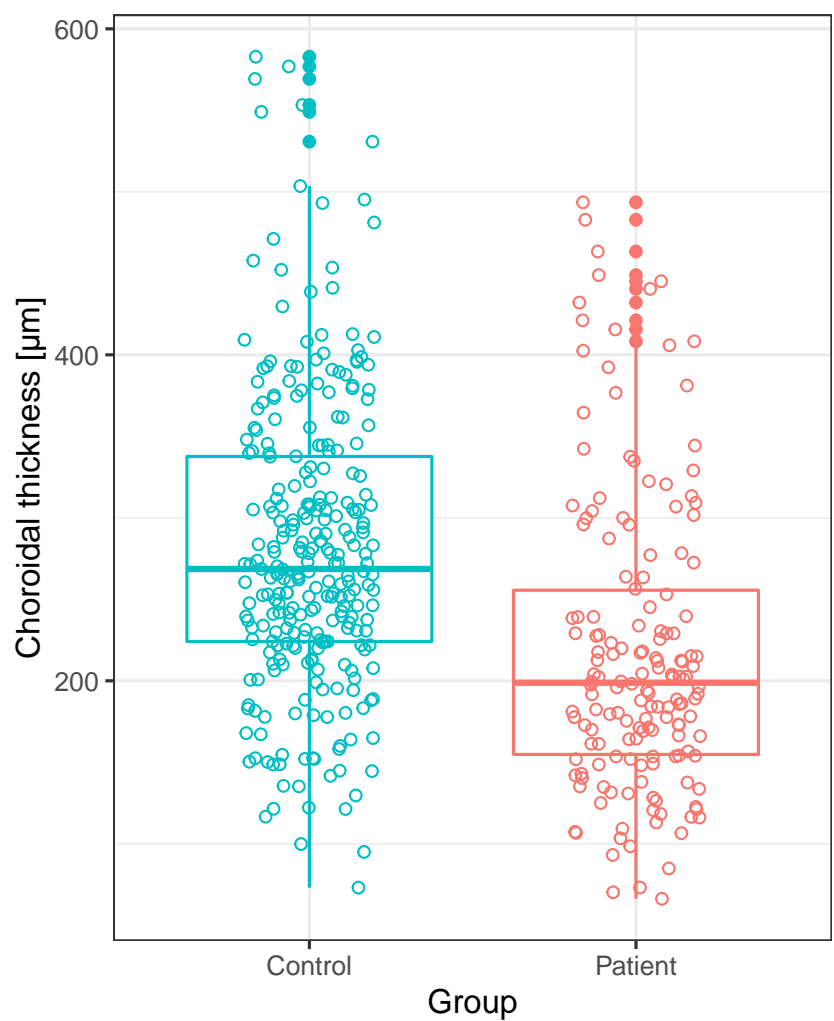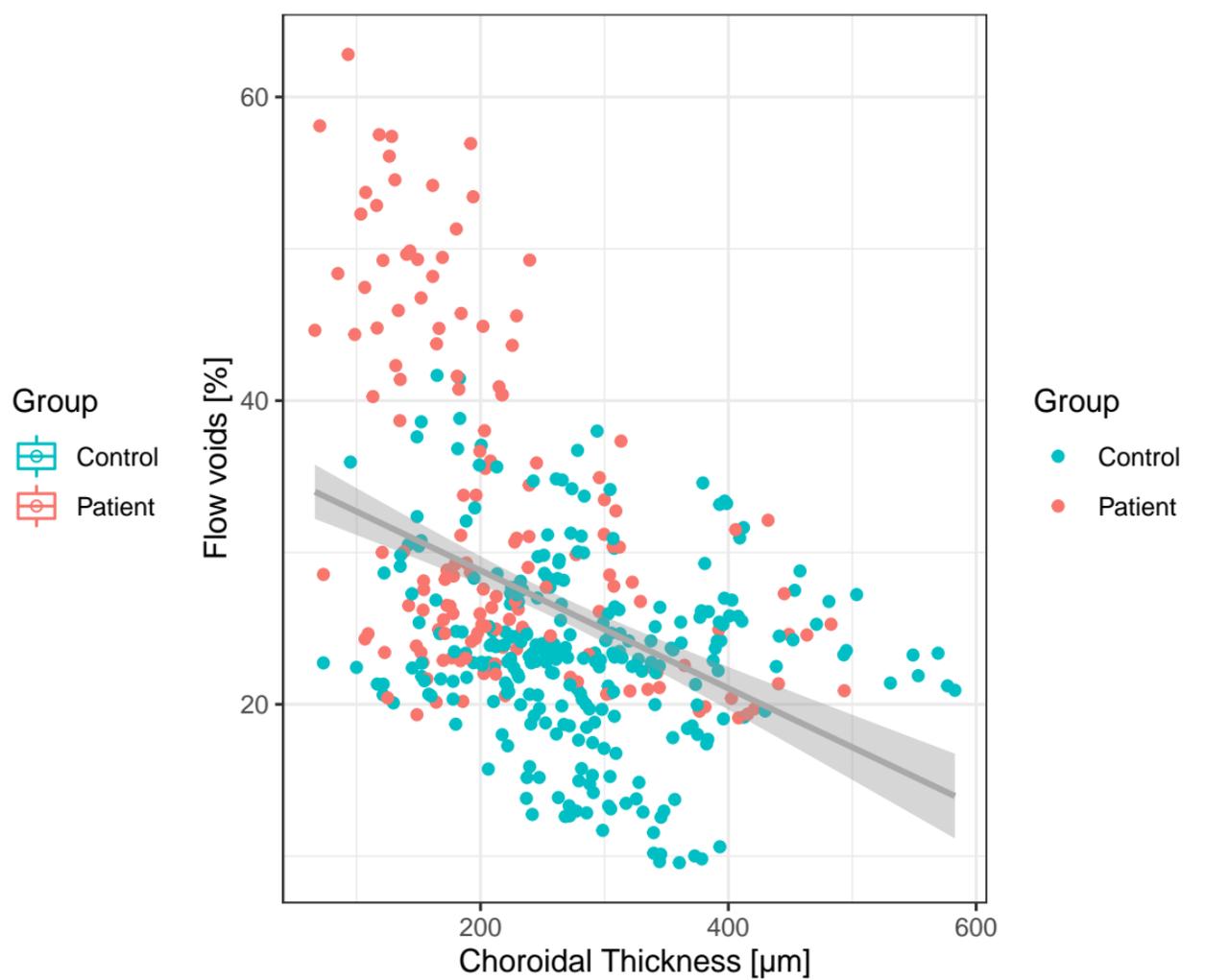

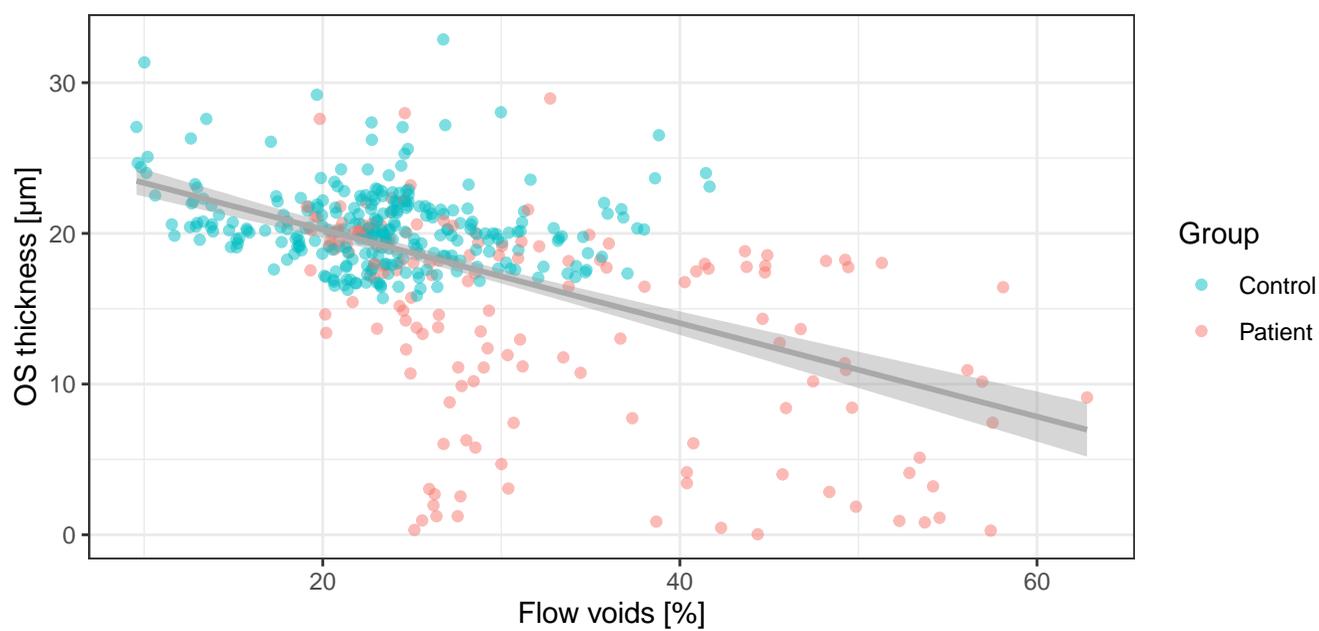

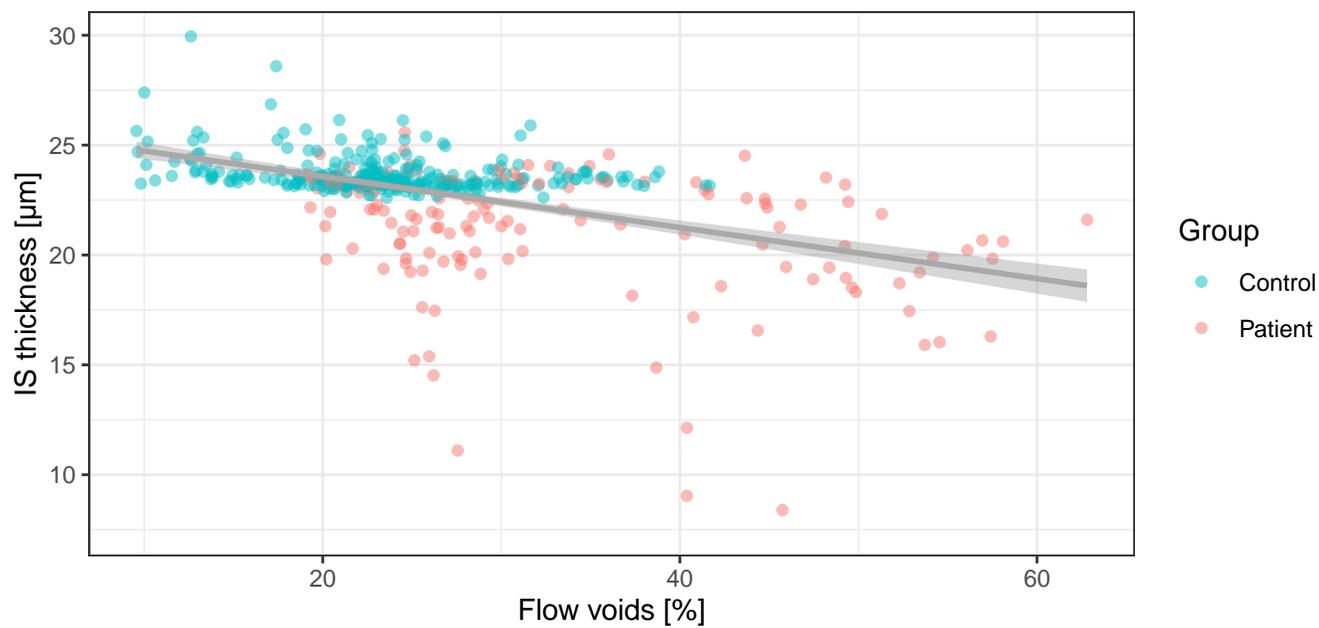

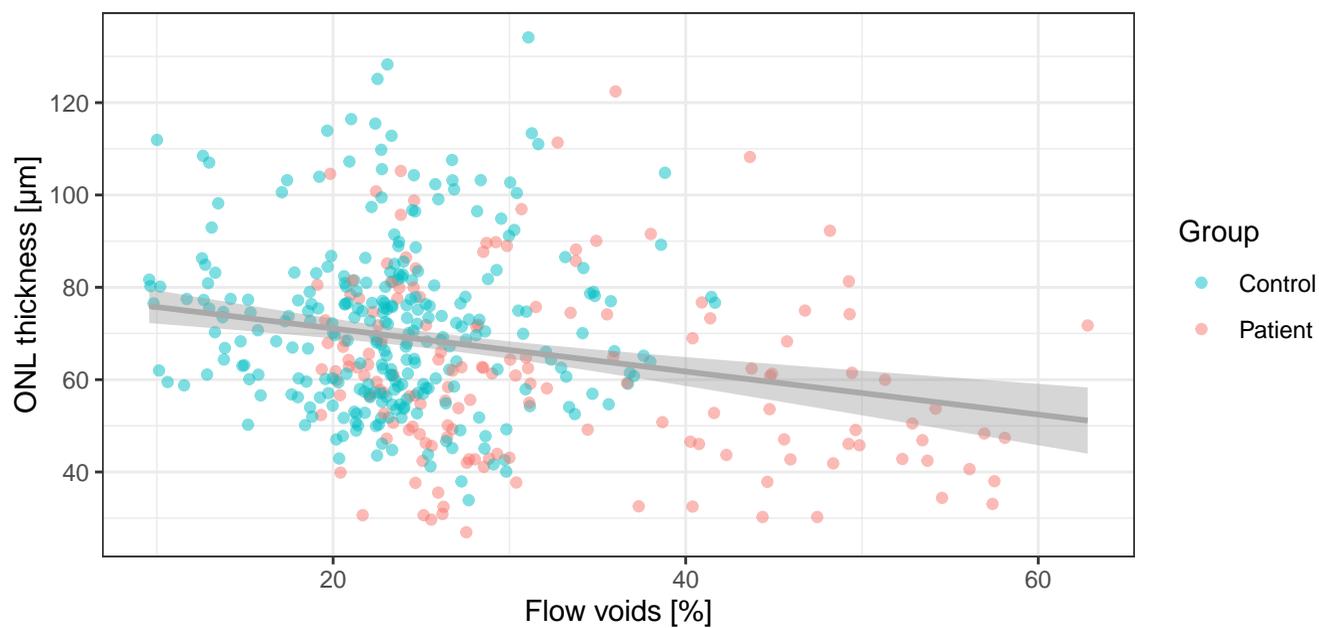

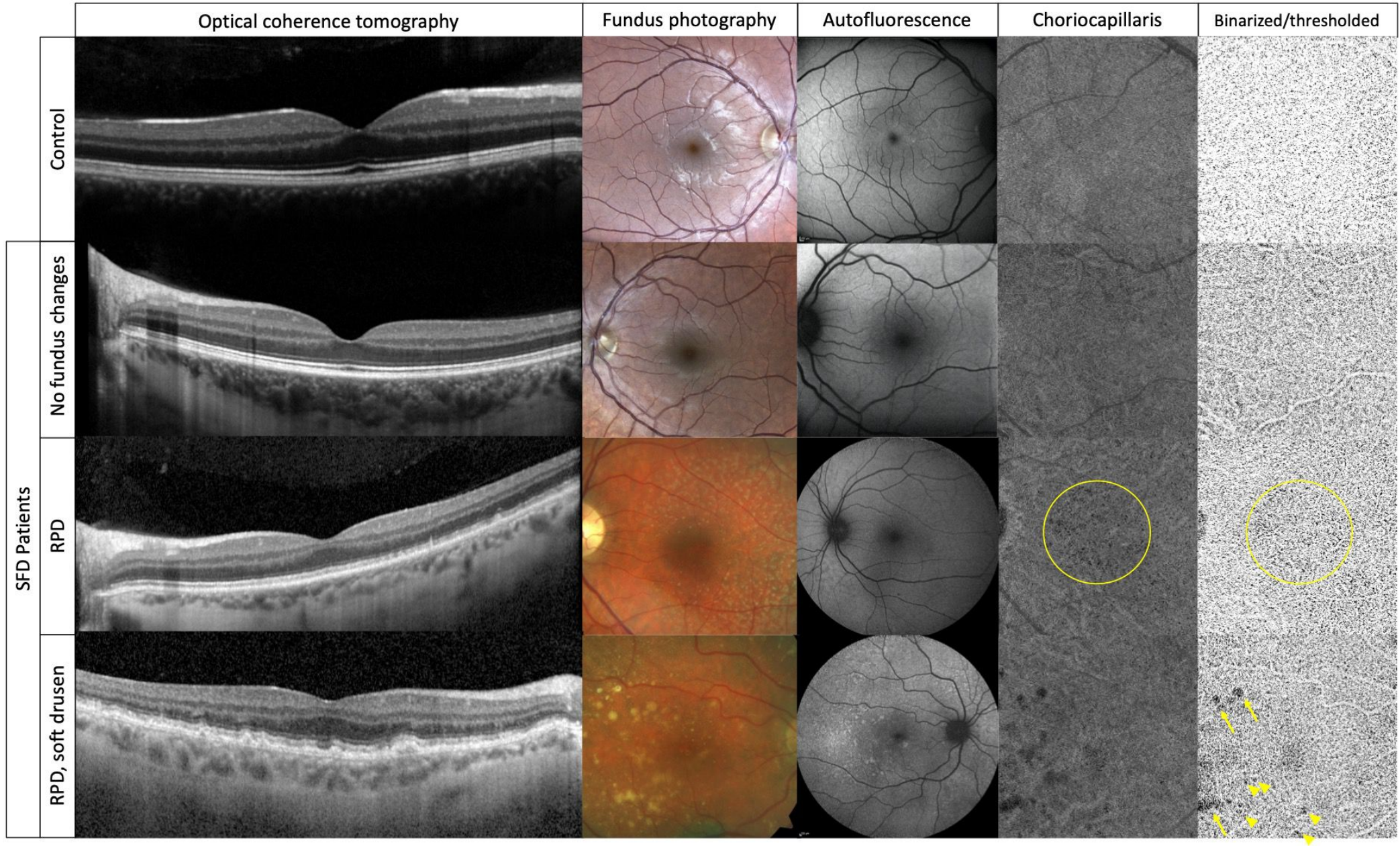

| | Optical coherence tomography | Fundus photography | Autofluorescence | Choriocapillaris | Binarized/thresholded |
|---|---|---|---|---|---|
| Control | | | | | |
| No fundus changes | | | | | |
| RPD | | | | | |
| RPD, soft drusen | | | | | |

SFD Patients